\input{aipcheck}

\documentclass[
    ,final            
  ]
  {aipproc}

\layoutstyle{6x9}

\usepackage{psfrag}

\begin{document}

\vspace*{-20mm}
\begin{flushright}
MPP-2009-151\\
\end{flushright}

\title{Tribrid Inflation in Supergravity
\footnote{Based on talks given by K.~D.\ at
Marcel Grossmann Meeting (Paris) and P.~M.~K. at SUSY09 (Boston) and 
PASCOS 2009 (Hamburg).
}}


\author{Stefan Antusch}{
  address={\it{
Max-Planck-Institut f\"ur Physik (Werner-Heisenberg-Institut),\\
F\"ohringer Ring 6,
80805 M\"unchen, Germany}}
}

\author{Koushik Dutta}{
}

\author{Philipp~M.~Kostka}{
}

\begin{abstract}
We propose a novel class of F-term hybrid inflation models in supergravity (SUGRA) where the $\eta$-problem is resolved using either a Heisenberg symmetry or a shift symmetry of the K\"ahler potential. In addition to the inflaton and the waterfall field, this class (referred to as tribrid inflation) contains a third 'driving' field which contributes the large vacuum energy during inflation by its F-term. 
In contrast to the ``standard'' hybrid scenario, it has several attractive features due to the property
of vanishing inflationary superpotential ($W_{inf}=0$) during inflation.  
While the symmetries of the K\"ahler potential ensure a flat inflaton potential at tree-level, quantum corrections induced by symmetry breaking terms in the superpotential generate a slope of the potential and lead to a spectral tilt consistent with recent WMAP observations.
\end{abstract}

\maketitle


\section{Introduction}
The state of the art in inflation model building 
offers a multitude of possibilities for realizing inflation~\cite{Lyth:1998xn}.
Among these many models, hybrid inflation is especially
promising to make a connection between the inflationary
paradigm and particle physics: The ``waterfall'' ending hybrid inflation 
may be associated with particle physics phase transitions such as the 
spontaneous breaking of the gauge group of a grand unified theory
or a flavor symmetry.  
In SUGRA, which provides a solution to the hierarchy problem
associated with such new physics at high energies, 
the $\eta$-problem is well known to put inflation 
models under considerable pressure~\cite{Copeland:1994vg}.  
In this talk, we propose a new variant of hybrid inflation models 
within SUGRA, where in addition to the inflaton and waterfall field, 
the model contains a 
third 'driving' field which contributes the large vacuum energy during 
inflation by its F-term. 
The new scenario, which we dub tribrid inflation, turns out to be 
particularly suitable for solving the $\eta$-problem
by symmetries in the K\"ahler potential. At the same time, it allows for
attractive connections to particle physics: The right-handed sneutrino, 
for example, provides an interesting inflaton candidate in
supersymmetric seesaw-extended versions of the Standard Model.

\section{Tribrid vs. Hybrid}
The general structure of the supersymmetric (SUSY) hybrid inflation
superpotential is given by~\cite{Copeland:1994vg, Dvali:1994ms}
\begin{equation}\label{hybrid}
W=\kappa\,\Phi\left(H^2-M^2\right)\,,
\end{equation}
where we denote the chiral superfield containing 
the slow-rolling inflaton as scalar component by $\Phi$
and the one containing the waterfall field by $H$. 
The F-term of the inflaton $\Phi$ contributes the
vacuum energy that drives inflation in this class of models
whereas inflation ends when the waterfall field acquires 
a vacuum expectation value (vev) $\langle H\rangle\sim M$.
For simplicity, we use only singlet fields.
Using gauge multiplets, one would substitute
$H^2\rightarrow H\bar{H}$, with $\bar{H}$ being 
another field in the conjugate representation.

Since $\Phi\ne 0$ during inflation in this class 
of models, it has the obvious properties
\begin{equation}\label{hybriddefinition}
W\ne0\,,\quad W_{\Phi}\ne0\,,
\end{equation}
where a lower index denotes derivative w.r.t.
the superfield.

As a new class of models, we propose the tribrid superpotential, given by
\begin{equation}\label{tribrid}
W=\kappa\,S\left(H^2-M^2\right)\,+\,g(\Phi,H)\,,
\end{equation}
which in addition to the previous two chiral superfields
in the hybrid superpotential, includes a so called 'driving'
field $S$. In contrast to the ``standard'' SUSY hybrid inflation models of the type in Eq.~\eqref{hybrid}, 
where the driving field is identical with the inflaton, 
in tribrid inflation each of the three main ingredients of the inflationary
model is distributed to a separate field\footnote{Hence the name tribrid inflation.}.
$S$ stays at zero during inflation and only contributes the
large vacuum energy by its F-term.
A large mass stabilizing S at zero is typically generated 
by SUGRA effects from generic non-minimal K\"ahler potentials.
$\Phi$ is the flat inflaton direction which slow-rolls and stabilizes
$H$ via the coupling $g(\Phi,H)$ until $\Phi$ reaches a critical
value and thus triggers the waterfall.
This ends inflation due to the fact that $H$ develops a
tachyonic mass squared and quickly falls towards 
the true vacuum $\langle H\rangle \sim M$.

The most simple renormalizable version would be
$g(\Phi,H)=\lambda\,\Phi\,H^2$. However, we favor
an operator of the form $\lambda\,\Phi^2\,H^2/M_*$ 
which could e.g. generate the right-handed (s)neutrino 
masses by the vev $\langle H\rangle \sim 10^{16}$ GeV in 
grand unified theories \cite{Antusch:2004hd}.  

Due to the fact that $S=H=0$ during inflation in the tribrid scenario defined in
Eq.~\eqref{tribrid}, one crucial difference to the hybrid scenario of Eq.~\eqref{hybrid} 
is a vanishing
superpotential and its derivative w.r.t. the inflaton
\begin{equation}\label{tribriddefinition}
W=0\,,\quad W_{\Phi}=0\,.
\end{equation}
In the next section, we discuss the implications of
these properties for possible solutions of the
$\eta$-problem.

\section{Advantages of the Tribrid}
With a general expansion of the K\"ahler potential
in terms of fields over some cutoff scale
and a suitable adjustment of the expansion parameters,
it is always possible to ``tune away'' the $\eta$-problem in both the 
``standard'' hybrid~\cite{BasteroGil:2006cm}
and the tribrid inflation scenarios~\cite{Antusch:2004hd}.

However, if one attempts to solve the $\eta$-problem by a fundamental symmetry
in the K\"ahler potential, this turns out to be extremely difficult
to achieve in ``standard'' hybrid-type models~\cite{Brax:2006ay, Davis:2008sa}.
The reason for this is that such symmetries typically lead to a tachyonic
direction in the potential which can only be stabilized at the cost of 
some extra complications, for instance by using the couplings to additional moduli 
fields which themselves have stabilization problems and induce dangerous couplings to
the inflaton via the SUGRA F-term scalar potential
\footnote{We use units where we set the reduced Planck scale
$M_{P}\sim 2.4\cdot 10^{18}$ GeV to one.}
\begin{equation} \label{Fterm}
V_{F}=e^K\left[K^{i\bar{j}}\,
D_{i}W\,D_{\bar{j}}\bar{W}
 - 3|W|^2\right]\,,
\end{equation}
where the derivative $D_i W\equiv W_i + W K_i$
has been introduced.

These problems do not arise if one combines the tribrid scenario in
Eq.~\eqref{tribrid} with a symmetry protecting the K\"ahler potential.
The main reason is that the conditions~\eqref{tribriddefinition}, in 
particular the vanishing of the inflationary superpotential $W_{inf}$ 
during inflation, avoids the appearance of tachyonic directions in
the potential. In addition, various potentially dangerous terms in the 
scalar potential, concerning for example couplings to moduli fields,
are automatically absent in tribrid inflation compared to the ``standard''
hybrid case.  
We have demonstrated this by constructing viable realizations of 
tribrid inflation in supergravity where the $\eta$-problem is solved 
naturally by either a Heisenberg symmetry or a shift symmetry of the K\"ahler 
potential \cite{Antusch:2008pn,Antusch:2009ef}.

\section{Heisenberg Symmetry Realization}
As a specific realization of the tribrid scenario, 
in Ref.~\cite{Antusch:2008pn} we have considered
the superpotential in Eq.~\eqref{tribrid} with
\begin{equation}\label{g}
g(\Phi,H)=\frac{\lambda}{M_*}\,\Phi^2\,H^2\,,
\end{equation}
in combination with a Heisenberg symmetry 
invariant K\"ahler potential of the form
\begin{equation}\label{HeisenbergKaehler}
K=\, |H|^2 + \left(1+\kappa_{S}\,|S|^2+\kappa_{\rho}\,\rho\,\right) |S|^2 + f(\rho)\,.
\end{equation}
The invariant combination under the non-compact 
Heisenberg group transformations is given by
$\rho=T+T^*-|\Phi|^2$.

The Heisenberg symmetry of the K\"ahler potential, or in other words the fact that $K$ depends
on $\rho$ only, together with the absence of kinetic mixing in the $(\rho,\Phi)$-basis, 
protects the potential Eq.~\eqref{Fterm} from containing 
SUGRA corrections to the mass of the inflaton $|\Phi|$.
We have shown that it is possible to stabilize the modulus
$\rho$ by the additional coupling $\kappa_{\rho}$ with the help
of the vacuum energy during inflation.
While the Heisenberg symmetry solves the $\eta$-problem
by keeping the tree-level potential exactly flat in $|\Phi|$-direction, 
one-loop corrections due to the Heisenberg symmetry 
breaking operator~\eqref{g}
with the waterfall sector fermions, scalars 
and pseudoscalars running in the loops
lift the flatness of the potential and generate the slope necessary for
slow-roll inflationary dynamics.

\section{Shift Symmetry Realization}
As another realization of the tribrid scenario, 
in Ref.~\cite{Antusch:2009ef} we have considered
the superpotential in Eq.~\eqref{tribrid},
again with the same function $g(\Phi,H)$ defined in Eq.~\eqref{g}.
This has been combined with a 
K\"ahler potential
\begin{equation}\label{kaehlerpotential}
 K=\,|H|^2+|S|^2+\frac{1}{2}\left(\Phi+\Phi^*\right)^2
+\frac{\kappa_{S}}{\Lambda^2}\,|S|^4
+\frac{\kappa_{\Phi}}{4\,\Lambda^2}\,\left(\Phi+\Phi^*\right)^4
+\frac{\kappa_{SH}}{\Lambda^2}\,|S|^2|H|^2\\
+\ldots\,,
\end{equation}
where the ellipsis symbolize all possible similar terms of the 
same order and
the suppressed higher order terms.
For fairly generic values of the couplings in the K\"ahler 
potential, it is possible to make all scalars in the theory
except for the inflaton
heavier than the Hubble scale during inflation and thus stable.

Due to the shift symmetry $\Phi\rightarrow\Phi+i\,\mu$ in the K\"ahler potential
we obtain a tree-level flat 
inflaton direction $\phi_{I}=\sqrt{2}\,Im (\Phi)$ and hence evade the $\eta$-problem.
Again, radiative corrections induced by the shift symmetry
breaking term in Eq.~\eqref{g} lift the flatness of the potential.
For sufficiently large values of the parameter $\kappa_{SH}$,
the loop-corrected potential can be of hilltop-form leading to a
stronger negative curvature of the inflaton potential and finally
allow for a reduced spectral 
index consistent with best-fit values to the WMAP 5 year data.

\section{Summary and Conclusions}
In summary, we have introduced a novel class of F-term inflation
models in SUGRA, which we dub tribrid inflation.
When combined with a Heisenberg or shift symmetry invariant 
K\"ahler potential, higher order operators from the SUGRA expansion
that give rise to the $\eta$-problem can be forbidden.
Due to the properties stated in Eq.~\eqref{tribriddefinition}, tribrid
inflation avoids stability problems which appear when ``standard'' hybrid inflation 
models are combined with fundamental symmetries in the K\"ahler potential.
Therefore, we conclude that tribrid inflation is tailor-made for solving 
the $\eta$-problem by symmetries in the K\"ahler potential.
Furthermore, it also allows for attractive connections to particle physics: 
The right-handed sneutrino, for example, provides an interesting 
inflaton candidate in tribrid inflation.


\begin{theacknowledgments}
The authors would like to thank Steve F. King and Mar Bastero-Gil for collaboration in
part of the works presented here. 
The authors acknowledge partial support by the DFG cluster of excellence 
``Origin and Structure of the Universe''.
\end{theacknowledgments}

\bibliographystyle{aipproc}   

\bibliography{sample}

\IfFileExists{\jobname.bbl}{}
 {\typeout{}
  \typeout{******************************************}
  \typeout{** Please run "bibtex \jobname" to optain}
  \typeout{** the bibliography and then re-run LaTeX}
  \typeout{** twice to fix the references!}
  \typeout{******************************************}
  \typeout{}
 }

\end{document}